# Structure-property-function relationships in triple-helical collagen hydrogels

Giuseppe Tronci,[1,2] Amanda Doyle,[1,2] Stephen J. Russell,[2] and David J. Wood[1]

[1] Biomaterials and Tissue Engineering Research Group, Leeds Dental Institute, University of Leeds, Leeds LS2 9LU, United Kingdom

[2] Nonwoven Research Group, Centre for Technical Textiles, University of Leeds, Leeds LS2 9JT, United Kingdom

## Abstract

In order to establish defined biomimetic systems, type I collagen was functionalised with 1,3-Phenylenediacetic acid (Ph) as aromatic, bifunctional segment. Following investigation on molecular organization and macroscopic properties, material functionalities, i.e. degradability and bioactivity, were addressed, aiming at elucidating the potential of this collagen system as mineralization template. Functionalised collagen hydrogels demonstrated a preserved triple helix conformation. Decreased swelling ratio and increased thermo-mechanical properties were observed in comparison to state-of-the-art carbodiimide (EDC)-crosslinked collagen controls. Ph-crosslinked samples displayed no optical damage and only a slight mass decrease (~ 4 wt.-%) following 1-week incubation in simulated body fluid (SBF), while nearly 50 wt.-% degradation was observed in EDC-crosslinked collagen. SEM/EDS revealed amorphous mineral deposition, whereby increased calcium phosphate ratio was suggested in hydrogels with increased Ph content. This investigation provides valuable insights for the synthesis of triple helical collagen materials with enhanced macroscopic properties and controlled degradation. In light of these features, this system will be applied for the design of tissue-like scaffolds for mineralized tissue formation.

## 1. Introduction

Collagen is the main protein of the human body, ruling structure, function and shape of biological tissues. Also in light of its unique molecular organization, collagen has been widely applied for the design of vascular grafts [1], fibrous materials for stem cell differentiation [2], biomimetic scaffolds for regenerative medicine [3], and tissue-like matrices for hard tissue repair [4]. However, collagen properties are challenging to control in physiological conditions, mainly because its hierarchical organization and chemical composition *in vivo* can only be partially reproduced *in vitro*. Functionalisation and crosslinking of collagen molecules, e.g. via carbodiimide [5,6], glutaraldehyde [7,8] or hexamethylene diisocyanate [9], have proved to enhance macroscopic properties in aqueous environment, although much is still left to do to establish biomimetic systems with defined structure-property-function relationships. Here, the design of type I collagen hydrogels was investigated via covalent lysine functionalisation with 1,3-Phenylenediacetic acid (Ph). It was hypothesized that incorporation of a stiff, aromatic segment among collagen molecules could offer a novel synthetic route to the formation of mechanically-relevant materials. Ph was selected as bifunctional segment, in order to promote crosslinking of distant collagen molecules, unlikely accomplished with current synthetic methods [1,5,6], so that triple helix conformation of collagen could be retained. The presence of an aromatic ring in the Ph was considered crucial to achieve controlled swelling and enhanced mechanical properties in resulting hydrogels, owing to the molecular stiffness of incorporated segment. In order to investigate the effectiveness of this synthetic approach, EDC treatment was selected as state-of-the-art reference method, since it has been shown to promote the formation of water-stable collagen materials with no residual toxicity, in contrast to aldehyde biomaterial fixation [1]. Formed hydrogels were investigated for molecular conformation, network architecture and hydrogel macroscopic properties, in comparison to EDC-crosslinked collagen. Furthermore, material degradability and bioactivity were addressed via incubation in SBF, in order to elucidate materials' potential for biomimetic mineralization.

## 2. Experimental details

1,3-Phenylenediacetic acid (Ph) was supplied by VWR International, all the other chemicals were purchased from Sigma-Aldrich. Type I collagen was isolated in-house from rat tail tendons [10]. Collagen was dissolved in 10 mM hydrochloric acid and functionalised with N-(3-Dimethylaminopropyl)-N′-ethylcarbodiimide hydrochloride (EDC)-activated Ph under gentle shaking at room temperature. EDC-crosslinked collagen was synthesized by mixing EDC with collagen solution, as previously reported [5]. Resulting hydrogels were washed with distilled water and dehydrated in aqueous solutions of increasing ethanol concentrations. Attenuated Total Reflectance Fourier Transform Infrared (ATR-FTIR) spectroscopy was carried out on dry samples using a Perkin-Elmer Spectrum BX spotlight spectrophotometer with diamond ATR attachment. 64 scans were averaged for each spectrum, using 4 cm$^{-1}$ resolution and 2 cm$^{-1}$ scanning interval. Degree of crosslinking ($C$) of collagen networks was determined by 2,4,6-trinitrobenzenesulfonic acid (TNBS) colorimetric assay (n=2) [11], as the molar ratio between functionalised and pristine, non functionalised lysines. Swelling tests (n=3) were carried out by incubating dry samples in 5 mL distilled water for 24 hours. Water-equilibrated samples were retrieved, paper-blotted and weighed. The weight-based swelling ratio ($SR$) was calculated as $SR=(m_s-m_d)/m_d \cdot 100$, where $m_s$ and $m_d$ are swollen and dry sample weights, respectively. Hydrogel discs (ø 0.8 cm, n=4) were compressed (Instron 5544 UTM) with a compression rate of 3 mm·min$^{-1}$. Differential Scanning Calorimetry (DSC) temperature scans were conducted on 10-140 °C temperature range with 10 °C·min$^{-1}$ heating rate (TA Instruments Thermal Analysis 2000 System and 910 Differential Scanning Calorimeter cell base). Mineralization experiment was carried out via 1-week sample incubation (n=4) at 25 °C SBF, with 0.7 sample weight/solution volume ratio [12]. Retrieved samples were rinsed with distilled water, dried, weighed and gold-coated for SEM/EDS (JEOL SM-35) analysis.

## 3. Discussion

### 3.1 Molecular organization of crosslinked collagen via ATR-FTIR spectroscopy

Triple helix collagen conformation is normally associated with three main amide bands, i.e. amide I at 1650 cm$^{-1}$, resulting from the stretching vibrations of peptide C=O groups; amide II absorbance at 1550 cm$^{-1}$, deriving from N–H bending and C–N stretching vibrations; and amide III band centered at 1240 cm$^{-1}$, assigned to the C–N stretching and N–H bending vibrations from amide linkages, as well as wagging vibrations of CH$_2$ groups in the glycine backbone and proline side chains [13]. Figure 1 indicates that the positions of these amide bands are maintained in Ph- as well as EDC-crosslinked networks. Furthermore, the FTIR absorption ratio of amide III to 1450 cm$^{-1}$ band was determined to be close to unity ($A_{III}/A_{1450}$ ~ 1.01-1.14) among the three samples, suggesting preserved integrity of triple helices [14] following functionalisation of native collagen.

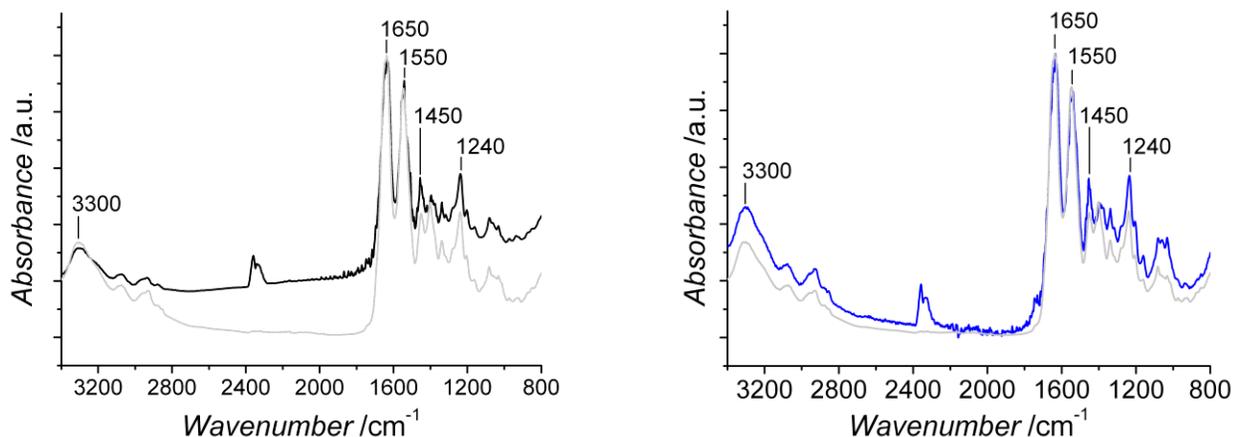

**Figure 1.** Exemplary FTIR spectra of Ph- (black line, left) and EDC-crosslinked (blue line, right) collagen. Native collagen (light gray line) spectrum is displayed for comparison in both plots.

### 3.2 Network architecture and macroscopic properties of collagen hydrogels

Network architecture was investigated both by quantifying the degree of crosslinking ($C$) via TNBS assay and by assessing the swelling ratio of resulting materials. As observed in Table I, collagen-Ph samples displayed a higher degree of crosslinking compared to state-of-the-art EDC-crosslinked collagen, despite having a very low Ph-collagen lysine molar ratio. A slight increase of Ph content (0.5 to 1.5 [COOH]/[Lys] ratio) in the crosslinking mixture led to nearly-complete functionalisation of collagen lysines (> 99 mol.-%). These results suggest that the employment of a bifunctional segment is likely to promote an increased yield of collagen functionalisation, since collagen molecules separated by a distance can be bridged [8]. On the other hand, EDC-mediated functionalisation results in the formation of zero-length net-points, whereby intramolecular crosslinks are likely established [5,8]. Consequently, steric hindrance effects almost certainly explain the decreased degree of crosslinking observed in collagen-EDC, compared to collagen-Ph, samples. Besides the degree of crosslinking, hydrogel swelling behaviour was also investigated; the swelling ratio of one sample, collagen-Ph1.5, was found to be significantly lower compared to all other EDC-based samples. This is supported by TNBS results, suggesting nearly complete functionalisation of collagen lysines for this composition. As for the other Ph-crosslinked samples (collagen-Ph0.5/1), swelling ratios were similar to each other, which is again supported by TNBS results, describing a similar degree of crosslinking.

In order to investigate whether variation of molecular parameters could induce changes in macroscopic properties, the thermo-mechanical properties of collagen hydrogels were addressed.

**Table I.** Degree of crosslinking (*C*), swelling ratio (*SR*) and denaturation temperature ($T_d$) of collagen hydrogels. [*]Sample are coded as 'Collagen-XXX-YY', where XXX indicates the type of system (either Ph or EDC-based), while YY identifies the molar ratio of either Ph carboxylic functions or EDC to collagen lysines.

| Sample ID[*] | *C* /mol.-% | *SR* /wt.-% | $T_d$ /°C |
|---|---|---|---|
| Collagen-Ph0.5 | 88 ± 3 | 1285 ± 450 | 80 |
| Collagen-Ph1 | 87 ± 14 | 1311 ± 757 | 80 |
| Collagen-Ph1.5 | > 99 | 823 ± 140 | 88 |
| Collagen-EDC10 | 25 ± 6 | 1392 ± 82 | 68 |
| Collagen-EDC20 | 37 ± 13 | 1595 ± 374 | 76 |
| Collagen-EDC30 | 34 ± 1 | 1373 ± 81 | 78 |
| Collagen-EDC40 | 68 ± 3 | 1374 ± 182 | 80 |
| Collagen-EDC60 | 60 ± 1 | 1106 ± 82 | 80 |

Collagen denaturation temperature ($T_d$) is related to the unfolding of collagen triple helices into randomly-coiled chains; it is therefore expected to be highly affected by the formation of a covalent network [2]. Table I describes hydrogel $T_d$ values as obtained by DSC; crosslinked samples show a denaturation temperature in the range of 68–88 °C, which is found to be higher than the denaturation temperature of native collagen ($T_d$ ~ 67 °C). Furthermore, variation of $T_d$ seems to be directly related to changes of crosslinking degree in the hydrogel network. These results give supporting evidence that covalent net-points were established during hydrogel formation, so that collagen triple helices were successfully retained and stabilized. It should be noted that Ph-crosslinked collagen revealed higher denaturation temperatures with respect to EDC- [6], glutaraldehyde- [7], and hexamethylene diisocyanate- [9] crosslinked collagen materials. This indicates that the incorporation of Ph as a stiff, aromatic segment superiorly stabilizes collagen molecules in comparison with current crosslinking methods.

Besides thermal analysis, mechanical properties of collagen-Ph hydrogels were measured by compression tests. Samples described *J*-shaped stress-compression curves (data not shown), similar to the case of native tissues. Here, shape recovery was observed following load removal up to nearly 50% compression, suggesting that the established covalent network successfully resulted in the formation of an elastic material, as observed in linear biopolymer networks [15]. On the other hand, EDC-crosslinked collagen showed minimal mechanical properties, whereby sample break was observed even after sample punching. Consequently, quantitative data of mechanical properties on collagen-EDC samples could not be acquired. These findings give further evidence that the collagen functionalisation with Ph is effective for the formation of collagen materials with enhanced mechanical properties. Compressive modulus ($E$: 28±10→35±9 kPa) and maximal stress of Ph-crosslinked samples ($\sigma_{max}$: 6±2→8±4 kPa) were measured in the kPa range, while compression at break ($\varepsilon_b$: 53±5–58±5 %) did not exceed 60% compression (Figure 2). The resulting compressive modulus was therefore measured to be almost 20 times higher compared to previously-reported collagen-based materials [8].

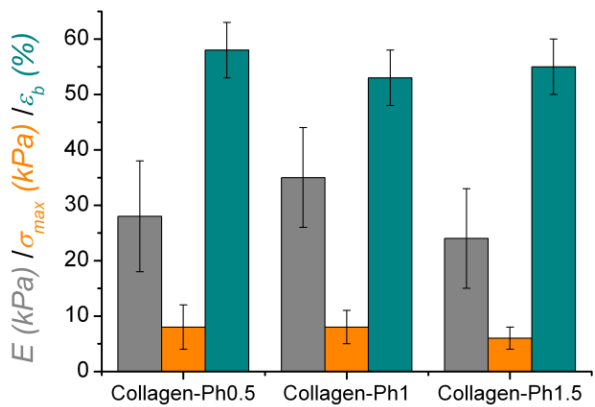

**Figure 2.** Compressive modulus ($E$), maximal compressive stress ($\sigma_{max}$), and compression at break ($\varepsilon_b$) of Ph-crosslinked collagen hydrogels (n=4).

At the same time, there was little variation in mechanical properties among the different compositions. Given the unique organization of collagen, the hierarchical level at which covalent crosslinks are introduced is crucial in order to study the influence of crosslinking on the mechanical properties of collagen. Olde Damink et al. observed no variation of mechanical properties in dermal sheep crosslinked collagen [6,7,9]. This was explained based on the fact that crosslinks were mainly introduced within rather than among collagen molecules. This hypothesis may be supported by above mechanical findings, although it is not in line with TNBS, swelling and thermal analysis data. Most likely, the variation of Ph feed ratio among the different compositions (0.5→1.5 [COOH]/[Lys] ratio) was probably too low to result in significant changes in mechanical properties. For these reasons, a wider range of Ph concentrations may be advantageous in order to establish hydrogels with varied mechanical properties. In that case, investigation via AFM will be crucial in order to explore the hierarchical levels at which covalent crosslinks are introduced.

**3.3 SBF incubation of collagen hydrogels**

SBF incubation is a well-known method to test a material's ability to form a hydroxycarbonate apatite (HCA) layer *in vitro* [12]. Collagen is known to trigger bone-like apatite deposition *in vivo* during bone formation [4], so it was of interest to investigate collagen hydrogel behaviour in SBF as an osteogenic-like medium. The mass change as well as the presence of calcium/phosphorous elements was therefore quantified in retrieved samples in order to (i) clarify any occurrence of degradation and (ii) determine the chemical composition of any potentially-nucleated phases.

A slight decrease in mass (averaged mass loss ~ 4 wt.-%) was observed in collagen-Ph samples, indicating minimal hydrolytic degradation had occurred. SEM on retrieved samples revealed nearly-intact material surfaces, confirming that a covalent network was still present at the molecular level. Among the different compositions, only one sample, collagen-Ph1, displayed a slight mass increase; observed differences in hydrolytic degradation may be related to a varied crosslinking/grafting ratio in the formed hydrogel networks. Networks with increased yield of grafting will likely degrade faster compared to networks with increased yield of crosslinking, since grafted molecules are expected to be cleaved more easily by water compared to crosslinked molecules. Consequently, variation of Ph feed ratio may affect the yield of crosslinking, so that grafted as well as crosslinked molecules may be formed above a specific Ph ratio threshold. A much higher mass loss (averaged mass loss ~ 53 wt.-%) was observed in EDC- compared to Ph-crosslinked collagen, which is in line with previous findings. Here, small micro-pores were observed on the retrieved sample surface, suggesting a surface, rather than bulk, erosion mechanism of hydrolytic degradation.

Besides degradation behaviour, SEM/EDS analyses were carried out to explore whether any mineral phase was nucleated following SBF incubation. Here, sample washing with water was crucial in order to remove superficial deposition of magnesium, sodium and chlorine ions, in agreement with the use of SBF. The presence of calcium and phosphorous elements was observed in all samples, although with a low Ca/P atomic ratio (0.84-1.41). This suggests that the mineral phase laid down on the material surface was most likely constituted of amorphous calcium phosphate, which may be expected due to the relatively short incubation time (1 week) at room instead of body temperature. Interestingly, sample collagen-Ph1.5 displayed increased Ca/P atomic ratio (Ca/P ~ 1.41) compared to the other samples, likely hinting at enhanced and selective nucleation of an apatite layer in hydrogels crosslinked with increased Ph feed ratio.

## 4. Conclusions

This study highlights the important role played by network molecular architecture on the macroscopic properties and functions of resulting collagen hydrogels. Following a bottom-up synthetic approach, functionalisation with a bifunctional, aromatic segment, successfully led to the establishment of a biomimetic system with preserved triple helix integrity and enhanced macroscopic properties, in comparison to state-of-the-art crosslinked collagen. Resulting hydrogels displayed minimal hydrolytic degradation following 1-week incubation in SBF, whereby nucleation of amorphous calcium phosphate phase was initiated.

## Acknowledgements

This work was funded through WELMEC, a Centre of Excellence in Medical Engineering funded by the Wellcome Trust and EPSRC, under grant number WT 088908/Z/09/Z. The authors would like to thank J. Hudson, W. Vickers and S. Finlay, for kind assistance with SEM/EDS, SBF preparation, and compression tests, respectively.